\documentclass[aps,prx,groupedaddress,twocolumn,showpacs,longbibliography,10pt]{revtex4-2}

\DeclareFontShape{OT1}{cmr}{mx}{n}{<->cmr10}{}

\fontseries{mx}\selectfont
\usepackage{color}
\usepackage{bm}
\usepackage{units}
\usepackage{bbm}

\usepackage[caption=false]{subfig}
\usepackage{graphicx}
\usepackage{dsfont}
\usepackage{amsmath}
\usepackage{amssymb}
\usepackage{array}
\usepackage{epstopdf}
\usepackage{enumerate}
\usepackage{youngtab}
\usepackage{tensor}
\usepackage{braket}
\usepackage[normalem]{ulem}
\usepackage{slashed}
\usepackage[aligntableaux=center]{ytableau}
\usepackage[utf8]{inputenc}
\usepackage[
      colorlinks=true,
      linkcolor=blue,
     urlcolor=blue,
      filecolor=black,
      citecolor=red,
      ]{hyperref}
\usepackage{cleveref}
\usepackage{braket}
\usepackage{mathtools}
\usepackage{rotating}
\usepackage{tabularx}
\newcolumntype{Y}{>{\centering\arraybackslash}X}
\newcolumntype{C}[1]{>{\centering\arraybackslash}p{#1}}

\usepackage{todonotes}
\usepackage{xcolor}
\definecolor{LightCyan}{rgb}{0.7,1,1}
\definecolor{Gray}{gray}{0.9}
\usepackage{xspace}

\begin{document}

\title{Magic Angles In Equal-Twist Trilayer Graphene}

\author{Fedor K. Popov and Grigory  Tarnopolsky}
\affiliation{Department of Physics, New York University, New York, NY 10003, USA}
\affiliation{Department of Physics, Carnegie Mellon University, Pittsburgh, PA 15213, USA}


\begin{abstract}
We consider a configuration of three stacked graphene monolayers 
with equal consecutive twist angles $\theta$. Remarkably, in the chiral limit when 
interlayer coupling terms between $\textrm{AA}$ sites of the moir\'{e} pattern are neglected we find four perfectly flat bands (for each valley) at a sequence of magic angles which are exactly equal to the twisted bilayer graphene (TBG) magic angles divided by $\sqrt{2}$. Therefore, the first magic angle for  equal-twist trilayer graphene (eTTG) in the chiral limit
is $\theta_{*} \approx 1.05^{\circ}/\sqrt{2} \approx 0.74^{\circ}$. We prove this relation analytically and show that the Bloch states of the eTTG's flat bands are non-linearly related  to those of TBG's. Additionally, we show that at the magic angles, the upper and lower bands must touch the four exactly flat bands at the Dirac point of the middle graphene layer. Finally, we explore the eTTG's spectrum away from the chiral limit through numerical analysis.

\end{abstract}


\maketitle
\nopagebreak

\section{Introduction}

The remarkable theoretical predictions \cite{doi:10.1073/pnas.1108174108, PhysRevB.82.121407} of the fascinating properties of twisted bilayer graphene (TBG) at a special ("magic") angle $\theta_{*}\approx 1.05^\circ$ and experimental realization of this configuration have uncovered an array of correlated phenomena, such as Mott insulating and superconducting phases \cite{CaoFatemiNature2018, CaoFatemiNature2, doi:10.1126/science.aav1910}. This tantalizing discovery has sparked an avalanche of further experimental and theoretical research \cite{PhysRevX.8.031089, PhysRevB.98.075109, PhysRevB.98.085435, doi:10.1073/pnas.1810947115, PhysRevB.98.085144, PhysRevB.98.195101, PhysRevLett.122.026801, PhysRevLett.122.086402, PhysRevB.98.035404, PhysRevB.98.045103, PhysRevLett.121.087001, PhysRevB.98.081102, PhysRevLett.121.257001, PhysRevB.99.075127,  PhysRevX.8.031088, Pizarro_2019, PhysRevX.8.031087, PhysRevB.98.241407, PhysRevX.8.041041, PhysRevB.98.235158, PhysRevB.98.220504,PhysRevB.99.144507, PhysRevB.106.235157, PhysRevLett.122.106405, PhysRevLett.123.036401, PhysRevB.99.035111, PhysRevB.99.195455}, 
aimed at gaining a deeper understanding of the underlying physics of these Van der Waals heterostructures.

The investigation of multiple graphene layer configurations, 
such as twisted trilayer graphene, is a natural generalization of the study of twisted bilayer graphene \cite{PhysRevB.100.085109, PhysRevLett.123.026402, CeaWaletGuinea2019, PhysRevLett.125.116404, PhysRevLett.123.026402, mao2023supermoire, lin2022energetic, ma2023doubled, PhysRevB.105.195422}. 
The multilayer systems possess greater number of parameters, which enhance their tunability. The initial theoretical investigation of twisted trilayer graphene (TTG) \cite{PhysRevB.100.085109, PhysRevLett.123.026402} unveiled a similar  flattening of electronic bands at various "magic"  angles, which ultimately led to experimental discovery of correlated phenomena and other intriguing physics \cite{PhysRevLett.127.166802, ParkCaoJarilloNature2021, doi:10.1126/science.abg0399,  LiuLiNaturePhys2022, doi:10.1126/science.abk1895, uri2023superconductivity}. The interacting effects  in such systems are under persistent   theoretical investigation \cite{PhysRevB.103.195411, PhysRevB.104.115167, PhysRevX.12.021018}.  
Other twisted graphene multilayer systems were discussed theoretically \cite{LeeKhalafShangNature2019, PhysRevLett.128.176404, ledwith2021tb, PhysRevLett.128.176403, ZhangXieWu2023, yang2023flat} and realized experimentally  \cite{LiuZeyuKhalafNature2020, CaoJarilloNature2020, ParkCaoJarilloNatureMat2022} where similar interacting effects were discovered.

\begin{figure}
\centering
\includegraphics[scale=0.12]{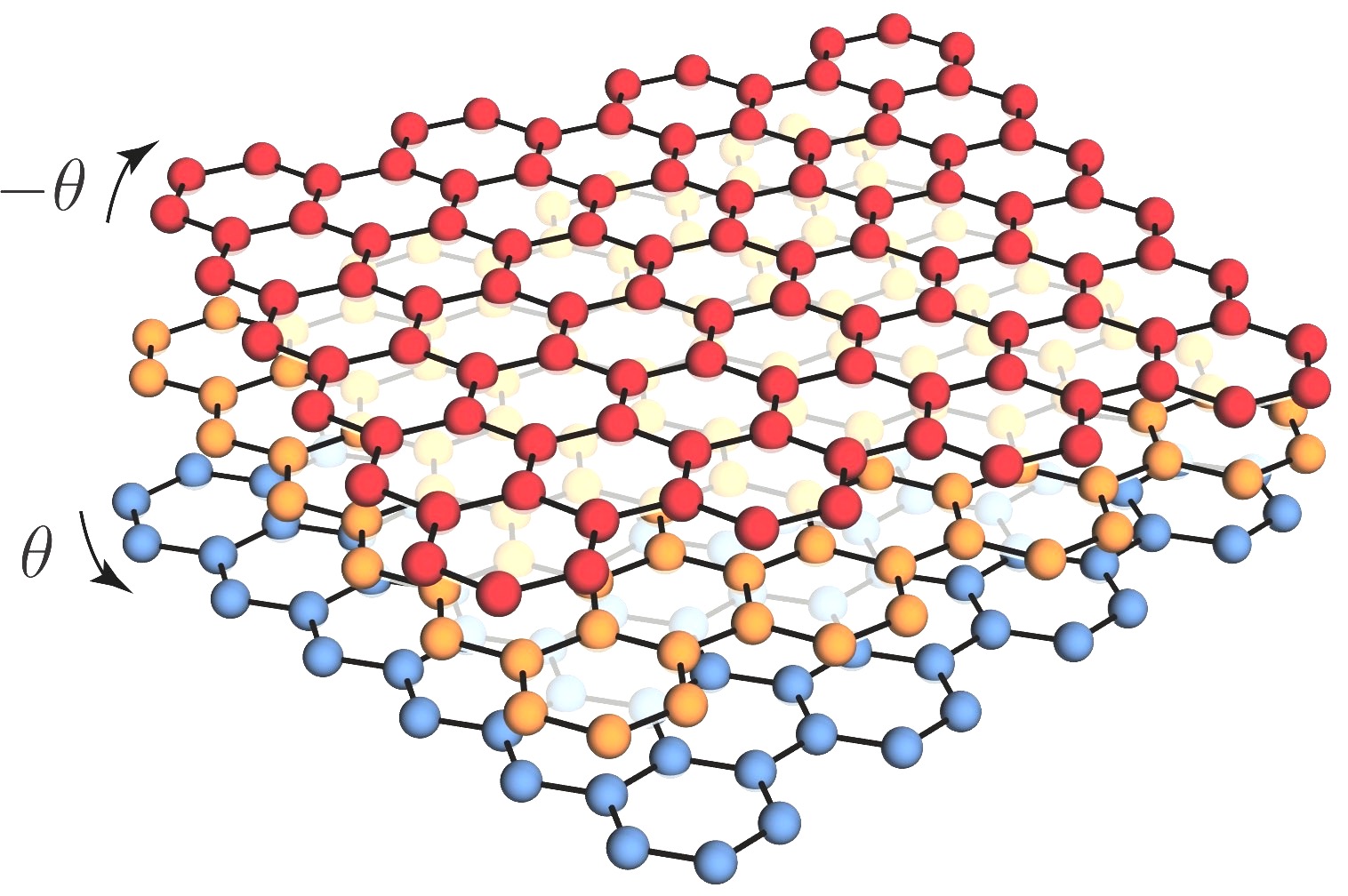}
\caption{A schematic illustration of the equal-twist trilayer graphene (eTTG):   the top and bottom layers are twisted by an angle $\theta$ in opposite directions relative to the middle layer. }
\label{fig:eTTG}
\end{figure}

In this letter, we focus on twisted trilayer graphene with equal small consecutive twist angles $\theta$. We refer to such a system as equal-twist trilayer graphene (eTTG), and it is schematically represented in Fig \ref{fig:eTTG}.
This particular twist configuration  of TTG has already been discussed in \cite{PhysRevLett.123.026402, mao2023supermoire, lin2022energetic, ma2023doubled, PhysRevB.105.195422}.

The main result of this article is exact relations between magic angles and perfectly flat bands wave functions of eTTG  and  TBG systems in the chiral limit (i.e. the limit when 
interlayer hopping terms between $\textrm{AA}$ sites of the moir\'{e} pattern are neglected). We show below that the magic angles of these systems are related as 
\begin{align}
\theta_{\textrm{eTTG}}= \theta_{\textrm{TBG}}/\sqrt{2}\,. \label{anglerel}
\end{align}
therefore the first magic angle of eTTG is $\theta_{*} \approx 1.05^{\circ}/\sqrt{2} \approx 0.74^{\circ}$. Remarkably this relation is exactly inverse of the relation found in \cite{PhysRevB.100.085109} for alternating-twist trilayer graphene (aTTG):
\begin{align}
\theta_{\textrm{aTTG}}= \theta_{\textrm{TBG}}\sqrt{2}\,. \label{aTTGrel}
\end{align}
In contrast to the aTTG case the relation (\ref{anglerel}) can be established only in the chiral limit. Moreover the relation between the Bloch states of the eTTG's exactly flat bands  and those of TBG's is valid only at the magic angles.

The paper is organized as follows. In Sec. \ref{sec:contmodel} we formulate continuum model for twisted trilayer graphene and then consider the case of equal-twist configuration of trilayer graphene (eTTG). 
We obtain the Hamiltonian for  such a configuration in the chiral limit. 
In Sec. \ref{sec:reltoTBG} we present our main result, namely the relation between magic angels and perfectly flat bands Bloch states of eTTG and TBG in the chiral limit. In Sec. \ref{sec:disc} we discuss theoretical and experimental challenges arising from investigation of  eTTG away from the chiral limit.

\section{Continuum model for Twisted Trilayer Graphene}
\label{sec:contmodel}

We consider a system of three stacked graphene monolayers, where each layer $\ell=1,2,3$ is rotated counterclockwise by an angle $\theta_{\ell}$ around  an atom site and then shifted by a vector $\textbf{d}_{\ell}$, so atoms in
each layer are parametrized by $\textbf{r}=R_{\theta_{\ell}}(\textbf{R}+\tau_{\alpha})+\textbf{d}_{\ell}$, where $R_{\theta}= e^{-i\theta \sigma_{y}}$ is the 
rotation matrix and $\textbf{R}$ and $\tau_{\alpha}$ run over the lattice and sub-lattice sites. The continuum model Hamiltonian for twisted trilayer graphene can be written as \cite{PhysRevB.100.085109}:
\begin{align*}
H= \left( \begin{array}{ccc}
    -iv_{F}  \bm{\sigma}_{\theta_{1}}\bm{\nabla} & T^{12}(\textbf{r}-\textbf{d}_{12}) &0 \\ 
    T^{12\dag}(\textbf{r}-\textbf{d}_{12}) & -iv_{F}  \bm{\sigma}_{\theta_{2}}\bm{\nabla} & T^{23}(\textbf{r}-\textbf{d}_{23}) \\ 
    0 & T^{23\dag}(\textbf{r}-\textbf{d}_{23}) & -iv_{F}  \bm{\sigma}_{\theta_{3}}\bm{\nabla} \\ 
  \end{array} \right)\,, \label{HamFullInitial}
\end{align*}
where $v_{F}\approx 10^{6}$m/s is the  monolayer graphene Fermi velocity, $\bm{\sigma}_{\theta}\equiv e^{i\frac{\theta}{2}\sigma_{z}} \bm{\sigma}e^{-i\frac{\theta}{2}\sigma_{z}}$, $ \bm{\sigma} = (\sigma_{x}, \sigma_{y})$ and $\textbf{d}_{\ell \ell'}=\frac{1}{2}(\textbf{d}_{\ell}+\textbf{d}_{\ell'} +i\cot(\theta_{\ell'\ell}/2)\sigma_{y}(\textbf{d}_{\ell}-\textbf{d}_{\ell'}))$ is the moir\'{e} pattern displacement vector. The moir\'{e} potential between adjacent layers $\ell$ and $\ell'$ is
\begin{align}
T^{\ell \ell'}(\textbf{r}) = \sum_{n=1}^{3}T^{\ell \ell'}_{n}e^{-i\textbf{q}_{n}^{\ell \ell'}\textbf{r}}\,, 
\end{align}
where $T^{\ell \ell'}_{n+1}=w_{\textrm{AA}}^{\ell \ell'}\sigma_{0}+w_{\textrm{AB}}^{\ell \ell'}(\sigma_{x} \cos n\phi +\sigma_{y}\sin n\phi)$ and 
\begin{align}
\textbf{q}_{1}^{\ell \ell'} = 2k_{D}\sin(\theta_{\ell'\ell}/2)R_{\phi_{\ell \ell'}}(0,-1), \;\; \textbf{q}^{\ell \ell'}_{2,3}=R_{\pm \phi}\textbf{q}^{\ell \ell'}_{1}
\end{align}
with $\theta_{\ell \ell'} =\theta_{\ell}-\theta_{\ell'}$,  $\phi_{\ell \ell'}=(\theta_{\ell}+\theta_{\ell'})/2$, $\phi=2\pi/3$, and $k_{D}=4\pi/3\sqrt{3}a$ is the Dirac momentum of the monolayer graphene with lattice constant $a = 1.42$\AA. The coupling between adjacent layers $\ell$ and $\ell'$ is characterized by two parameters $w_{\textrm{AA}}^{\ell \ell'}$ and $w_{\textrm{AB}}^{\ell \ell'}$ representing intra- and intersub-lattice couplings. The chiral limit corresponds to 
 $w_{\textrm{AA}}^{\ell \ell'}=0$.

\begin{figure}
\centering
\includegraphics[scale=0.7]{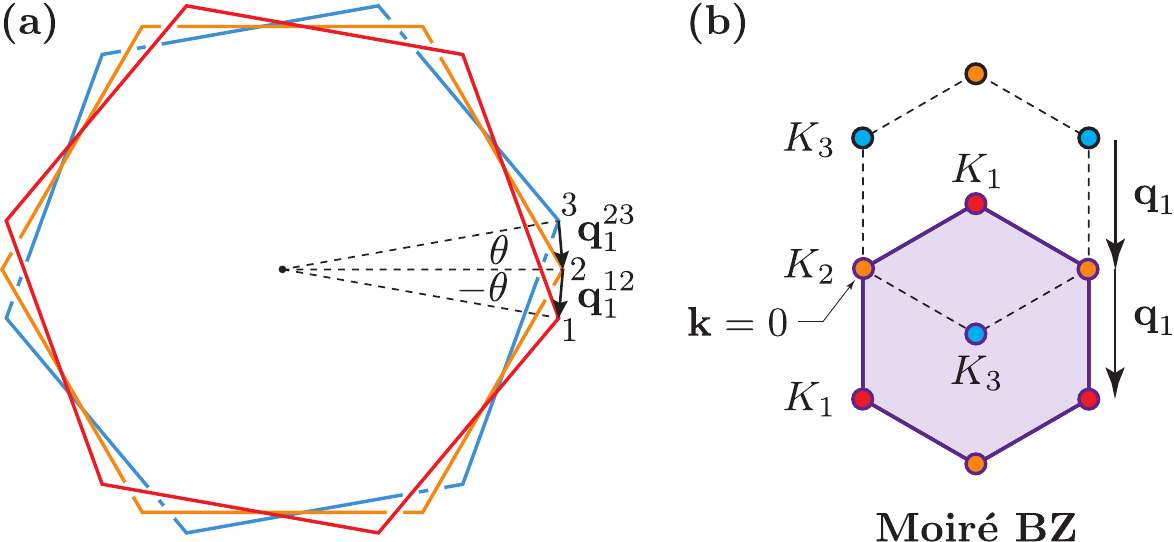}
\caption{(a) Original Brillouin zones of three  graphene layers with their Dirac points $K_{1}$, $K_{2}$ and $K_{3}$. Three layers are consecutively twisted by the same angle $\theta$, so $|\textbf{q}_{1}^{12}| =|\textbf{q}_{1}^{23}|$.   (b) Moir\'{e} Brillouin zone for  eTTG. We neglect a relative rotation between vectors $\textbf{q}_{1}^{12}$ and $\textbf{q}_{1}^{23}$  and denote them by $\textbf{q}_{1}$. The wave vector $\textbf{k}$ is zero at the Dirac point $K_{2}$.}
\label{fig:TLGstruct}
\end{figure}

In this letter we consider only a trilayer configuration with two equal consecutive twist angles $\theta$ thus we take $\theta_{1}=-\theta$, $\theta_{2}=0$ and $\theta_{3}=\theta$ (Magic angles in the chiral limit of a general TTG configuration with $\theta_{12}/\theta_{23}=p/q$, where $p$ and $q$ are coprime integers are discussed elsewhere \cite{PopovTarnTopub}, see also \cite{PhysRevLett.123.026402}).
Moreover we assume that there is no displacement between layers $\textbf{d}_{\ell \ell'}=0$ (for the aTTG case it was shown in \cite{CarrLiZiyan2020} that this is energetically favorable stacking configuration). Finally for a small angle $\theta$ we set $\phi_{\ell \ell'} = 0$ leading to $\textbf{q}_1^{12} = \textbf{q}_1^{23} = \textbf{q}_1$ 
 \cite{PhysRevLett.123.026402}.  Thus we obtain the following Hamiltonian for  eTTG:
\begin{gather}
H_{\textrm{eTTG}} = \begin{pmatrix}
 -i v_F \bm{\sigma}_{-\theta}\bm{\nabla} & T(\mathbf{r}) & 0\\
 T^\dagger(\mathbf{r}) & -i v_F \bm{\sigma} \bm{\nabla} & T(\mathbf{r}) \\
 0 & T^\dagger(\mathbf{r}) &- i v_F \bm{\sigma}_{\theta}\bm{\nabla}
\end{pmatrix}, \label{eq:Hamtril}
\end{gather} 
where we also assumed that the coupling parameters $w_{\textrm{AA}}$ and $w_{\textrm{AB}}$ do not depend on layers.  The moir\'{e} Brillouin zone (mBZ) for this Hamiltonian is depicted in Fig. \ref{fig:TLGstruct}. The  reciprocal moir\'{e} lattice is generated by vectors $\mathbf{b}_{1,2} = \mathbf{q}_{2,3}-\mathbf{q}_2$. In the coordinate space the eTTG configuration forms a single moir\'{e} lattice that is spanned by the lattice vectors $\mathbf{a}_{1,2} = (4\pi/3 k_\theta) (\pm \sqrt{3}/2, 1/2)$, with $k_{\theta}=2k_{D}\sin(\theta/2)$, and we neglect the effects of moir\'{e} of moir\'{e} lattice \cite{mao2023supermoire}.  It is useful to introduce complex coordinates  $z, \bar{z}=\mathbf{r}_x \pm  i \mathbf{r}_y$ in real space and $k, \bar{k} = \mathbf{k}_1 \pm  i \mathbf{k}_2$ in momentum space.
The Hamiltonian (\ref{eq:Hamtril}) acts on a spinor $\Phi(\textbf{r})=(\phi_{1},\eta_{1}, \phi_{2},\eta_{2},\phi_{3},\eta_{3})$, where the indices $1,2,3$ represent the graphene layer.

For a small twist angle $\theta$ we can neglect the phase factors in the Pauli matrices $\bm{\sigma}_{\pm \theta} \to \bm{\sigma}$.  Introducing the dimensionless variable $\alpha = w_{\textrm{AB}}/(v_{F} k_{\theta})$, and
writing the Hamiltonian \eqref{eq:Hamtril} in the sublattice basis $\Phi(\textbf{r})=(\phi_{1}, \phi_{2}, \phi_{3}, \eta_{1},\eta_{2},\eta_{3})$ we obtain
\begin{align}
\mathcal{H}_{\textrm{eTTG}} = \begin{pmatrix}
 \mathcal{M}(\textbf{r}) & \mathcal{D}^{\dag}_{3}(\textbf{r})\\
\mathcal{D}_{3}(\textbf{r}) &  \mathcal{M}(\textbf{r})
\end{pmatrix}\,,
 \label{eq:eTTG}
\end{align}
where we have rescaled coordinates $\textbf{r} \to k_{\theta}\textbf{r}$ and the Hamiltonian, so the energies of (\ref{eq:eTTG}) are measured in units of $v_{F}k_{\theta}$. 
The operators $\mathcal{D}_3$ and $\mathcal{M}$ are 
\begin{align}
&\mathcal{D}_3(\textbf{r}) = \left(  \begin{array}{ccc}
    -2i\bar{\partial} & \alpha U(\textbf{r}) & 0 \\ 
    \alpha U(-\textbf{r}) & -2i\bar{\partial} & \alpha U(\textbf{r}) \\ 
    0 & \alpha U(-\textbf{r}) & -2i\bar{\partial} \\  
  \end{array}\right)\,, \notag\\
&\mathcal{M}(\mathbf{r}) = \frac{w_{\textrm{AA}}}{w_{\textrm{AB}}} \left(  \begin{array}{ccc}
    0 &  U_{0}(\textbf{r}) & 0 \\ 
     U_{0}(-\textbf{r}) & 0 &  U_{0}(\textbf{r}) \\ 
    0 &  U_{0}(-\textbf{r}) &0 \\  
  \end{array}\right)\,,  \label{eq:D3andM}
\end{align}
where the potentials $U(\mathbf{r}) = \sum^3_{n=1} \omega^{n-1} e^{-i \textbf{q}_n \textbf{r}}$ and 
$U_{0}(\mathbf{r}) = \sum^3_{n=1} e^{-i \textbf{q}_n \textbf{r}}$, $\omega = e^{i\phi}$,  $\textbf{q}_{n} = R_{(n-1)\phi}(0,-1)$
and we introduced  derivatives $\partial, \bar{\partial} = \frac{1}{2}(\partial_{x} \mp i \partial_{y})$.
The Bloch states $\Phi_{\textbf{k}}(\textbf{r})=(\phi_{\textbf{k}}(\textbf{r}), \eta_{\textbf{k}}(\textbf{r}))$ of  \eqref{eq:eTTG}  are parametrized by the wave vector $\textbf{k}$ from mBZ and satisfy the following boundary conditions
\begin{gather}
\Phi_\mathbf{k}(\mathbf{r}+\mathbf{a}_{1,2}) = e^{i \mathbf{k}\, \mathbf{a}_{1,2}}  U^{(3)}_\phi \Phi_\mathbf{k}(\mathbf{r})\,, \label{BCTTG}
\end{gather}
where the matrix $U^{(3)}_{\phi} = \mathds{1}_{\textrm{AB}}\otimes \operatorname{diag}\left(\omega,1,\omega^{*}\right)$.

\begin{table}
 \caption{Comparison between magic angles for TBG and eTTG in the chiral limit  ($w_{\textrm{AA}}/w_{\textrm{AB}}=0$).}
 \label{tab:angles}
\begin{center}
 \begin{tabular}{c | c| c| c |c } 
 \hline
 \hline
  & $\alpha_1$ & $\alpha_2$ & $\alpha_3$ & $\alpha_4$  \; \\ [0.5ex] 
 \hline
 TBG & $0.586$ & $2.221$ & $3.75$ & $5.276$   \\ 
 \hline
eTTG& $0.829$ &   $3.141$ & $5.30$ & $7.461$ \\
 \hline
\end{tabular}
\end{center}
\end{table}

\section{Relation to the twisted bilayer graphene}
\label{sec:reltoTBG}
In this section we show that  eTTG Hamiltonian (\ref{eq:eTTG}) has an infinite series of magic angles and exactly flat bands in the chiral limit $w_{\textrm{AA}} =0$.  Moreover the magic angles and the Bloch states of the eTTG's flat bands are related to those of TBG's. Namely, if TBG has two exactly flat bands at the magic angle $\theta_{\textrm{TBG}}$ then  eTTG must have four exactly flat bands at the magic angle at $\theta_{\textrm{eTTG}}=\theta_{\textrm{TBG}}/\sqrt{2}$. 

The Hamiltonian for TBG in the chiral limit has the following form  in the sublattice  basis \cite{PhysRevLett.122.106405}: 
\begin{align}
\mathcal{H}_{\textrm{TBG}} = \begin{pmatrix}
0 & \mathcal{D}^{\dag}(\mathbf{r})\\
\mathcal{D}(\mathbf{r}) & 0
\end{pmatrix}, \,\,
\mathcal{D}(\mathbf{r}) = \left(  \begin{array}{ccc}
    -2i\bar{\partial} & \alpha U(\mathbf{r}) \\ 
    \alpha  U(-\mathbf{r}) & -2i\bar{\partial} \\ 
  \end{array}\right)\,. \label{eq:TBG}
\end{align}
The Bloch's wave functions $\Psi_{\textbf{k}}(\textbf{r})=(\psi_{\textbf{k}}(\textbf{r}), \chi_{\textbf{k}}(\textbf{r}))$ satisfy the
boundary conditions
\begin{gather}
\Psi_\mathbf{k}(\mathbf{r}+\mathbf{a}_{1,2}) = e^{i \mathbf{k}\, \mathbf{a}_{1,2}}  U^{(2)}_{\phi} \Psi_\mathbf{k}(\mathbf{r})\,, \label{BCTBG}
\end{gather}
where $U^{(2)}_\phi = \mathds{1}_{\textrm{AB}}\otimes \operatorname{diag}\left(1,\omega^{*}\right)$. This Hamiltonian has two exactly flat bands at zero energy over the entire mBZ:
\begin{gather}
\mathcal{H}_{\textrm{TBG}} \Psi_\mathbf{k} (\textbf{r})= \varepsilon_{0}(\textbf{k})\Psi_\mathbf{k}(\textbf{r}), \quad  \varepsilon_{0}(\textbf{k}) =0 \notag
\end{gather}
at the infinite series of magic angles  $\alpha=0.586, 2.221, \dots$. These flat bands are 
formed by the Bloch states $\Psi_{\mathbf{k}} = \left(\psi_{\mathbf{k}},0\right)$ and $\hat{\Psi}_{\mathbf{k}} = \left(0, \chi_{\mathbf{k}}\right)$, where the functions $\psi_{\mathbf{k}}$ satisfy the equation
\begin{gather}
\mathcal{D}(\mathbf{r}) \psi_{\mathbf{k}}(\mathbf{r}) = 0, \quad \psi_{\mathbf{k}} (\mathbf{r}+\mathbf{a}_{1,2}) = e^{i \mathbf{k} \mathbf{a}_{1,2} } U_\phi \psi_{\mathbf{k}}(\mathbf{r}),    \label{aceq}
\end{gather}
with $U_\phi = \operatorname{diag}\left(1,\omega^{*}\right)$ and the functions $\chi_{\mathbf{k}}$ satisfy $\mathcal{D}^{\dag}\chi_{\textbf{k}} =0$ with the same boundary conditions.
 
\begin{figure}[t!]
\centering
\includegraphics[width = 3.4in]{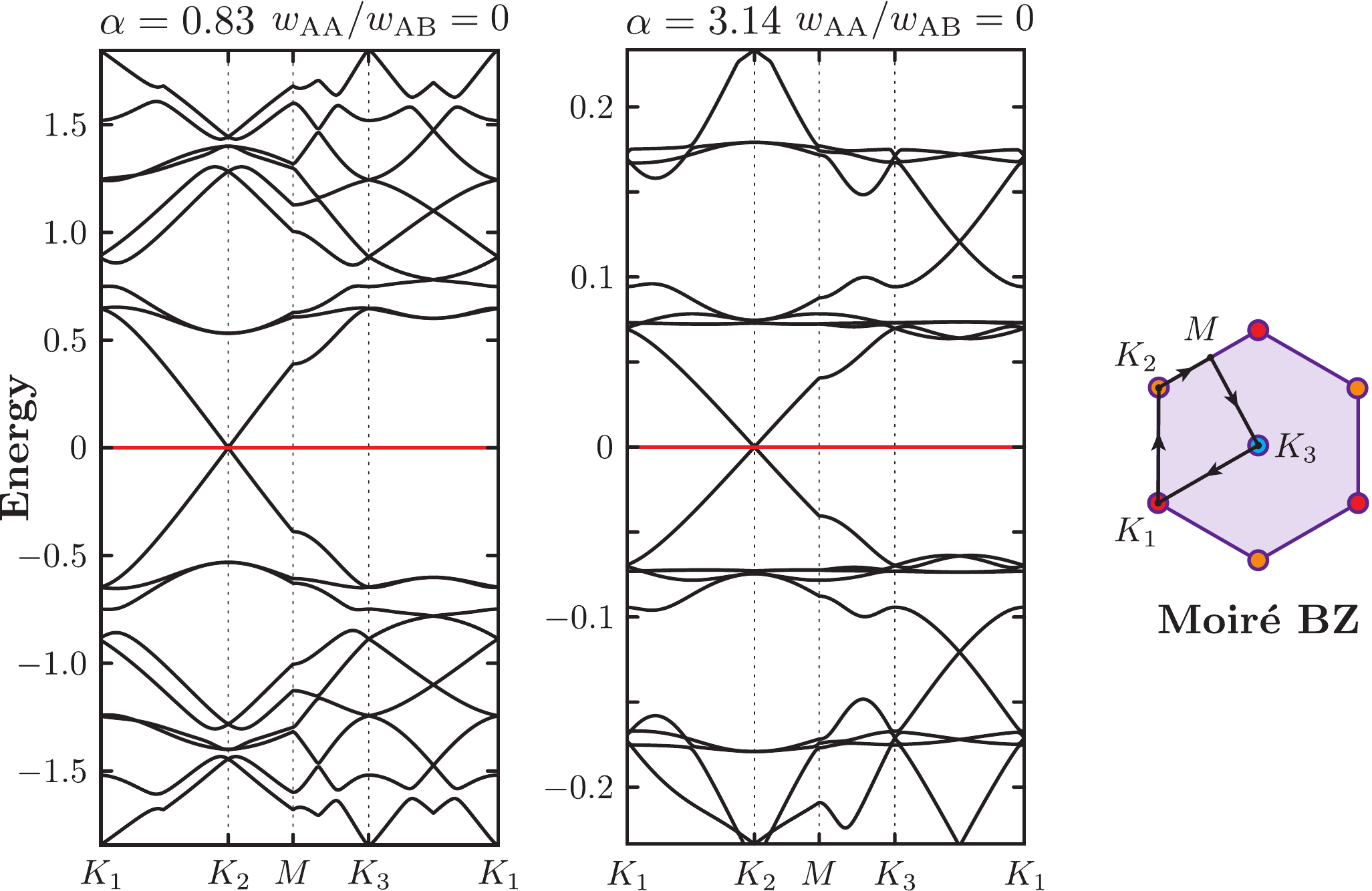}
\caption{Spectrum of the equal-twist trilayer graphene Hamiltonian (\ref{eq:eTTG}) in the chiral limit ($w_{\textrm{AA}}/w_{\textrm{AB}}=0$) at the first two magic angles $\alpha=0.83$ and $\alpha=3.14$. There are four exactly flat bands at zero energy (highlighted in red).}
\label{fig:sp1}
\end{figure}

Let us take two wave functions $\psi_{\mathbf{k}}$ and $\psi_{\mathbf{k}'}$ at  wave vectors $\mathbf{k}$ and $\mathbf{k}'$ of the mBZ (we notice that the eTTG and TBG Hamiltonians (\ref{eq:eTTG}) and (\ref{eq:TBG}) have identical mBZ). These wave functions are two-component spinors: $\psi_{\mathbf{k}} = \left(\psi_{\mathbf{k}1},\psi_{\mathbf{k}2}\right)$ and $\psi_{\mathbf{k}'} = \left(\psi_{\mathbf{k}'1},\psi_{\mathbf{k}'2}\right)$ and  we can construct the following three-component wave function using their components:
\begin{align}
&\phi_{\mathbf{k} + \mathbf{k}' + \mathbf{q}_1}(\mathbf{r}) = \psi_\mathbf{k}(\mathbf{r})\times \psi_{\mathbf{k}'}(\mathbf{r})   \label{eq:mytrans} \\ 
&\qquad \equiv\begin{pmatrix}
\psi_{\mathbf{k}1}(\mathbf{r})\psi_{\mathbf{k}'1}(\mathbf{r})\\
\frac{1}{\sqrt{2}} \left(\psi_{\mathbf{k}1}(\mathbf{r})\psi_{\mathbf{k}'2}(\mathbf{r}) + \psi_{\mathbf{k}2}(\mathbf{r})\psi_{\mathbf{k}'1}(\mathbf{r})\right)\\
\psi_{\mathbf{k}2}(\mathbf{r})\psi_{\mathbf{k}'2}(\mathbf{r})
\end{pmatrix}\notag\,.
\end{align}
It is possible to check that this wave function satisfies the zero energy equation of  eTTG:
\begin{align}
\mathcal{D}_3(\mathbf{r}) \phi_{\mathbf{k} + \mathbf{k}' + \mathbf{q}_1} = 0 \,, \label{D3zero}
\end{align}
and the boundary conditions (\ref{BCTTG}), provided we use the following relation between
the parameters $\alpha$ (twist angles) of eTTG and TBG:  
\begin{align}
\alpha_{\textrm{eTTG}} = \sqrt{2} \alpha_{\textrm{TBG}}\,, \quad  
(\theta_{\textrm{eTTG}} = \theta_{\textrm{TBG}}/\sqrt{2})\,. 
\end{align}
Hence we constructed zero energy Bloch states $\Phi_{\textbf{k} + \textbf{k}' + \textbf{q}_1}=(\phi_{\textbf{k} + \textbf{k}' + \textbf{q}_1},0)$ of the Hamiltonian $\mathcal{H}_{\textrm{eTTG}}$ in the chiral limit $w_{\textrm{AA}} 
=0$.  Similarly we can construct  zero energy Bloch states $\hat{\Phi}_{\textbf{k} + \textbf{k}' + \textbf{q}_1}=(0,\eta_{\textbf{k} + \textbf{k}' + \textbf{q}_1})$ using the functions $\chi_{\textbf{k}}$ and  $\chi_{\textbf{k}'}$ of TBG. The spectrum of eTTG in the chiral limit at the first two magic angles is depicted in Fig. \ref{fig:sp1} and we listed the first four magic angles in TBG and eTTG in the Table \ref{tab:angles}.
Below we write explicit expressions for the eTTG flat bands Bloch states and explain why there are four exactly flat bands in this case.

It was derived in \cite{PhysRevLett.122.106405} that the wave functions $\psi_{\textbf{k}}(\mathbf{r})$ in (\ref{aceq}) have the form 
   \begin{align}
&\psi_{\textbf{k}}(\textbf{r}) =f_{\textbf{k}}(z) \psi_{K}(\textbf{r})\,, \notag\\
&f_{\textbf{k}}(z) = e^{i \frac{\textbf{k}\textbf{a}_{1}}{a_{1}}z}\frac{\vartheta_{1}((z-(1-ik)z_{0})/a_{1}|\omega)}{\vartheta_{1}((z-z_{0})/a_{1}|\omega)} \,,\label{TBGwf}
  \end{align} 
where $a_{1,2}=(\textbf{a}_{1,2})_{x} + i(\textbf{a}_{1,2})_{y}$, $z_{0}=\frac{1}{3}(a_{1}-a_{2})$
and $\psi_{K}= (\psi_{K,1},\psi_{K,2})$ is a solution of (\ref{aceq}) at the Dirac point $K$, which corresponds to 
$\textbf{k}=0$ (this solution exists for an arbitrary twist angle). The theta-function 
$\vartheta_{1}(z|\tau)$ is defined as 
     \begin{align}
\vartheta_{1}(z|\tau) = \sum_{n=-\infty}^{+\infty} e^{i\pi \tau(n+\frac{1}{2})^{2}} e^{2\pi i (z-\frac{1}{2})(n+\frac{1}{2})}\,,
  \end{align} 
and has zeros at $z= m+ n \tau$.  At the magic angles, the function $\psi_{K}(\textbf{r})$ has a zero at the point $\textbf{r}_{0}=\frac{1}{3}(\textbf{a}_{1}-\textbf{a}_{2})$. This zero cancels with the zero of the theta-function 
in (\ref{TBGwf}), making $\psi_\mathbf{k}(\mathbf{r})$ finite.  
 The function $\psi_{\textbf{k}}(\textbf{r})$ has zero at $z=(1-ik)z_{0}$ due to the theta-function in the numerator. 

Using these results for TBG we obtain for the eTTG zero energy wave functions 
\begin{align}
\phi_{\textbf{k} }(\mathbf{r}) = f_{\textbf{k}'}(z)f_{\textbf{k} - \textbf{k}'-\textbf{q}_{1}}(z) \phi_{K_{1}}(\textbf{r})\,, \label{phiffpsi}
 \end{align} 
where $ \phi_{K_{1}} = ((\psi_{K,1})^{2}, \sqrt{2} \psi_{K,1} \psi_{K,2}, (\psi_{K,2})^{2})$ is the zero energy solution at the Dirac point $K_{1}$ of eTTG (corresponds to $\textbf{k}=\textbf{q}_{1}$). 
The functions (\ref{phiffpsi}) have two zeros (or a double zero) in the moir\'{e} lattice unit cell and thus there are only two linearly independent solutions (\ref{phiffpsi}) at each point of the mBZ  \cite{PhysRevB.31.2529}. 
There is a freedom to  choose a basis of two linearly independent solutions $\phi_{\textbf{k}}^{(1)}$ and $\phi_{\textbf{k}}^{(2)}$ and one possible choice is: 
\begin{align}
&\phi_{\textbf{k}}^{(1)}(\textbf{r}) =f_{\textbf{k}}(z) f_{-\textbf{q}_{1}}(z)  \phi_{K_1}(\mathbf{r})\,, \notag\\
&\phi_{\textbf{k}}^{(2)}(\textbf{r}) = f_{\textbf{k}+\textbf{q}_{1}}(z) f_{-2\textbf{q}_{1}}(z)  \phi_{K_1}(\mathbf{r})\,. \label{BasisSetPhi}
 \end{align} 
We notice that these functions are not necessarily orthogonal. It is possible to construct an orthogonal set following \cite{PhysRevB.31.2529}. Hence  the functions $\Phi_{\textbf{k}}=(\phi_{\textbf{k}},0)$ and $\hat{\Phi}_{\textbf{k}}=(0, \eta_{\textbf{k}})$ comprise four exactly flat bands.

Now we show that at the magic angles of the chiral eTTG the upper and lower bands touch the four exactly flat bands at the Dirac point  $K_{2}$, as can be seen in Fig. \ref{fig:sp1}.  The emergence of two additional zero modes $\tilde{\Phi}_{K_{2}}=(\tilde{\phi}_{K_{2}},0)$ and $\tilde{\Phi}_{K_{2}}=(0,\tilde{\eta}_{K_{2}})$ at the magic angles is related to existence of unphysical singular solutions $\tilde{\psi}_{\textbf{k}}(\textbf{r})$ and  $\tilde{\chi}_{\textbf{k}}(\textbf{r})$ of the chiral TBG Hamiltonian at the magic angles  \cite{PhysRevB.103.155150}. As was shown in \cite{PhysRevB.103.155150}  the equation (\ref{aceq}) apart from the regular solution $\psi_{\textbf{k}}(\textbf{r})$ in (\ref{TBGwf}),  admits a singular solution $\tilde{\psi}_{\textbf{k}}(\textbf{r})$ which has a pole instead of a zero. Since this solution is singular it never appears in the spectrum of TBG. But in the case of eTTG the relation (\ref{eq:mytrans}) allows to 
construct a non-singular wave function $\tilde{\phi}_{K_{2}}(\textbf{r})$ by multiplying components of the   
function $\tilde{\psi}_{\textbf{k}}$ with the pole by the components of the function $\psi_{\textbf{k}'}$ with zero, such that  the pole and  zero are at the same point of the moir\'{e} unit cell and cancel each other. This is possible provided $\textbf{k}+\textbf{k}'+\textbf{q}_{1} =0$.  In spite of infinitely many combinations $\textbf{k}$ and $\textbf{k}'$ satisfying this constraint, there is only a single linearly independent wave function $\tilde{\phi}_{K_{2}}(\textbf{r})$ and for concreteness we choose $\textbf{k}=0$ and 
$\textbf{k}'=-\textbf{q}_{1}$, so we can write 
\begin{gather}
\tilde{\phi}_{K_{2}}(\mathbf{r}) = \tilde{\psi}_K(\mathbf{r}) \times \psi_{K'}(\mathbf{r})\,,
\end{gather}
where $\tilde{\psi}_{K}$ and $\psi_{K'}$ are the singular and regular solutions of the equation (\ref{aceq})
at the TBG Dirac points $K$ ($\textbf{k}=0$) and $K'$ ($\textbf{k}=-\textbf{q}_{1}$) respectively. A similar construction applies to the function $\tilde{\eta}_{K_{2}}(\textbf{r})$.

\begin{figure}[t!]
\centering
\includegraphics[width = 3.4in]{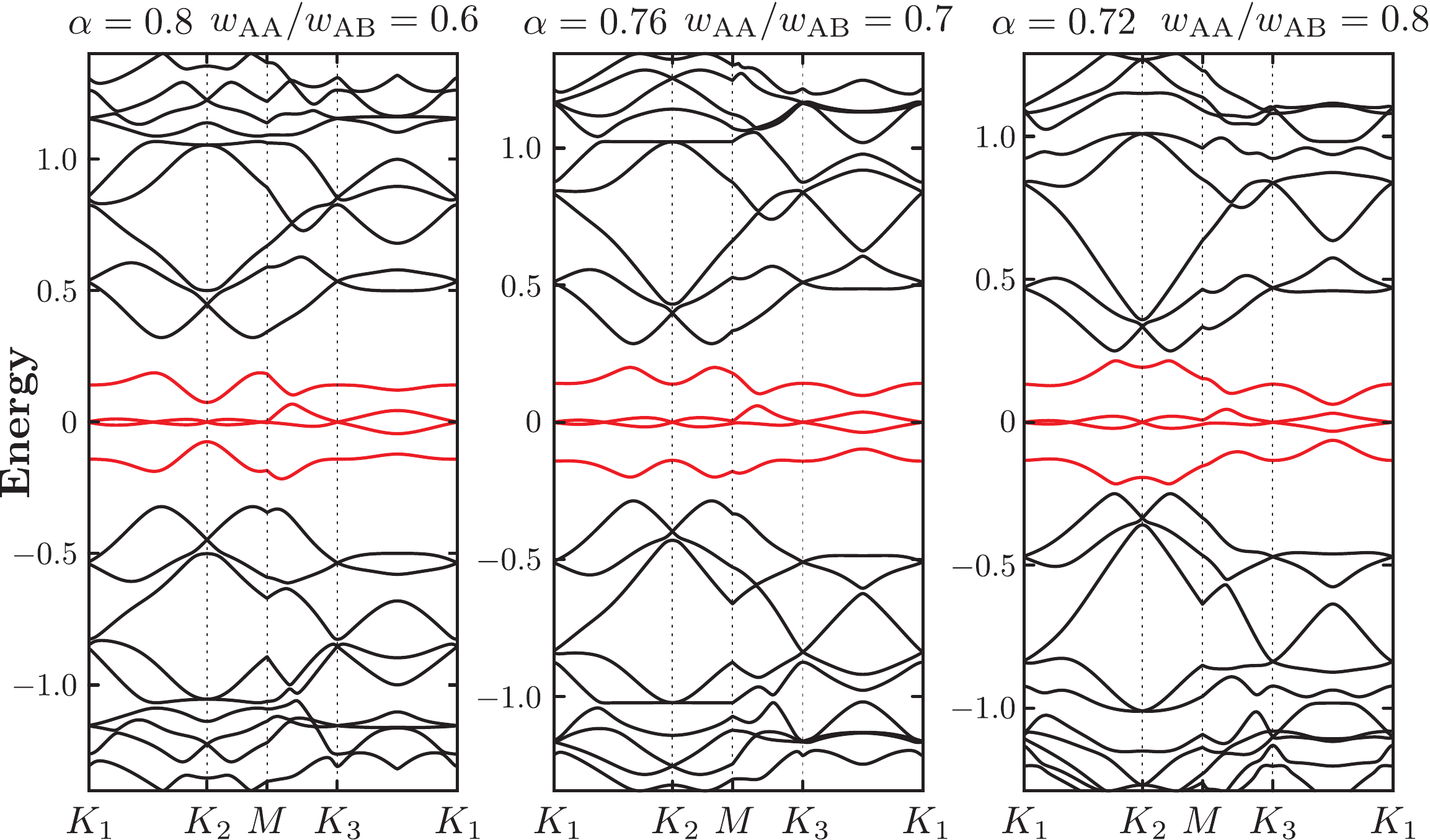}
\caption{Spectrum of the equal-twist trilayer graphene Hamiltonian (\ref{eq:eTTG}) for $w_{\textrm{AA}}/w_{\textrm{AB}}=0.6, 0.7, 0.8$ at the twist angles $\alpha =0.8, 0.76, 0.72$ close to the first magic angle $\alpha =0.83$. For each ratio $w_{\textrm{AA}}/w_{\textrm{AB}}$  we chose a twist angle such that it minimizes the band width of the first two lowest energy bands. The four lowest energy bands are highlighted in red. These bands are exactly flat  in the chiral limit ($w_{\textrm{AA}}/w_{\textrm{AB}}=0$) at the magic angle.}
\label{fig:sp2}
\end{figure}

Finally in Fig. \ref{fig:sp2} we plot a spectrum of the eTTG Hamiltonian (\ref{eq:eTTG}) away from the chiral limit. 
We see that the four lowest energy bands are sensitive to the coupling  parameter $w_{\textrm{AA}}$. Nevertheless 
the first two lowest energy bands remain relatively flat in a small range of twist angles close to the first eTTG magic angle $\alpha=0.83$.

\bigskip

\section{Discussion} 
\label{sec:disc}

In conclusion,  we related magic angles and the Bloch states of the exactly flat bands of eTTG to those of  TBG in the chiral limit. The electronic band structure of TBG  has  been extensively studied  \cite{PhysRevLett.122.106405, PhysRevB.103.155150,  PhysRevResearch.2.023237, PhysRevResearch.3.023155, PhysRevLett.127.246403, sheffer2022symmetries, parhizkar2023generic}.  We remark that the relation between the alternating-twist trilayer graphene (aTTG) and TBG  found  in \cite{PhysRevB.100.085109} is a linear algebraic relation between the aTTG and TBG Hamiltonians and works for any twist angles $\alpha$ and ratio of the coupling parameters $w_{\textrm{AA}}/w_{\textrm{AB}}$. In contrast, the relation  between the eTTG and TBG zero energy Bloch states is non-linear and is valid only at the magic angles or Dirac points. 
In general the rest of the spectrum of  eTTG and TBG at the magic angles is different.

Finally we remark that the current experimental works did not investigate eTTG at the range of angles close to the first magic angle 
$\theta_{*} \approx 0.74^{\circ}$ predicted in this letter.  Although the lowest energy bands of eTTG near the magic angle lose their perfect flatness away from the chiral limit, it is still possible that these bands "remember" the holomorphic structure of the Bloch states (\ref{BasisSetPhi}) in the chiral limit, similarly to the case of  TBG \cite{TarnopoUnpub} and this could potentially pave the way for exciting new discoveries.

\section*{Acknowledgments}  

We are grateful to A. Devarakonda, S. Chatterjee, I. R. Klebanov and V. Kozii for useful discussions. We would like to thank I. R. Klebanov for  valuable comments on the draft.
F.K.P. is currently a Simons Junior Fellow at New York University and supported
by a grant 855325FP from the Simons Foundation.

\bibliographystyle{ieeetr} 
\bibliography{Trilayer}

\end{document}